\documentclass[twocolumn, superscriptaddress,  floatfix, unsortedaddress, aps, prl, showpacs, amssymb,amsmath]{revtex4}
\usepackage{graphics}
\usepackage{color}
\usepackage{epsfig}
\usepackage{epstopdf}

\newcommand{\new}[1] {#1}

\begin{document}

\title{Stress Relaxation in Entangled Polymer Melts}

\author{\new{Ji-Xuan Hou}}
\affiliation{Laboratoire de Physique  and Centre Blaise Pascal of the \'Ecole Normale Sup\'erieure de Lyon, Universit\'e de Lyon, CNRS UMR 5672,\\
46, all\'ee d'Italie,
69364 Lyon cedex 07, France}
\author{\new{Carsten Svaneborg}}
\affiliation{Department of Chemistry and Interdisciplinary Nanoscience Center (iNANO), University of Aarhus, Langelandsgade 140, DK-8000 Aarhus C, Denmark}
\author{Ralf Everaers}
\affiliation{Laboratoire de Physique  and Centre Blaise Pascal of the \'Ecole Normale Sup\'erieure de Lyon, Universit\'e de Lyon, CNRS UMR 5672,\\
46, all\'ee d'Italie, 69364 Lyon cedex 07, France}
\author{Gary S. Grest}
\affiliation{Sandia National Laboratories, Albuquerque, NM 87185, USA}

\begin{abstract}
We present an extensive set of simulation results for the stress relaxation in
equilibrium and step-strained bead-spring polymer melts. The data allow us to explore
the chain dynamics and the shear relaxation modulus, $G(t)$, into the plateau regime for chains
with $Z=40$ entanglements
and into the terminal relaxation regime for $Z=10$.
Using the known (Rouse) mobility of unentangled chains and
the melt entanglement length determined via the primitive path analysis
of the microscopic topological state of our systems, we have performed parameter-free tests of several different tube models.
\new{We find excellent agreement for the Likhtman-McLeish theory using the double reptation approximation for constraint release, if we remove the contribution of high-frequency modes to
contour length fluctuations of the primitive chain.}
\end{abstract}


\maketitle

High molecular weight polymeric
liquids display remarkable viscoelastic properties \cite{Ferry_80,Bird_77}.
Contrary to glassy systems, their macroscopic relaxation times
are not due to slow dynamics on the monomer scale, but arise
from the chain connectivity and the restriction that the backbones of
polymer chains cannot cross while undergoing Brownian motion.
Modern theories of polymer dynamics~\cite{doi86, mcleish02}
describe the universal aspects of the viscoelastic behavior based on
the idea that molecular entanglements confine individual polymers to
tube-like regions in space \cite{Edwards67,deGennes71}. 
Forty years of research  have led to a complex relaxation scenario based
on a combination of local Rouse dynamics, reptation, contour length fluctuations, and
constraint release~\cite{mcleish02}. The development and validation
of a quantitative, microscopic theory crucially depends on the availability
of experimental and simulation data for model systems.

Entangled polymers are studied experimentally using rheology~\cite{Ferry_80,Bird_77,BaillyPRL06},
dielectric spectroscopy~\cite{Watanabe}, small-angle neutron scattering~\cite{ewen97, Wischnewski_prl_02}, and
nuclear magnetic resonance~\cite{CallaghanSamulskiMM97,KSaalwaechter}.
Computer simulations~\cite{kremer90,KroegerHessPRL00,pant95,AuhlEveraersJCP} offer some advantages in the preparation of well-defined model systems and
the simultaneous access to macroscopic behavior and microscopic
structure and dynamics. In particular, the recently developed
primitive path analysis (PPA)~\cite{RE_PPA_sci_04,PPA2,Kroeger_cpc_05,ShanbhagLarson_prl_05,ZhouLarsonMM2005,Tzoumanekas_mm_06,Uchida_jcp_08,HoyPRE09}
reveals the experimentally inaccessible {\em mesoscopic}
structures and relaxation processes described by the tube model
and allows {\em parameter-free} comparisons between theoretical predictions and data.
However, the long relaxation times pose a particular
challenge to computational techniques. Here we present simulation results for
model polymer melts in equilibrium and after a rapid, volume-conserving
uni-axial elongation, where we have been able to follow the full relaxation dynamics
deep into the entangled regime. The data allow us to perform the first parameter-free test
of the predictions of tube models for dynamical properties, \new{to pinpoint a problem in the current theoretical description, and to validate a suitable modification.}

Our numerical results are based on extensive molecular dynamics (MD) simulations
of bead-spring polymer melts \cite{kremer90}.  Each
chain is represented as a sequence of beads connected by finite-extensible, non-linear
(FENE) springs and interacting via the repulsive part
of the Lennard-Jones 12-6 potential (LJ). The energy
scale is set by  the strength of the LJ interaction, $\epsilon$, while
the distance scale is set by the monomer size, $\sigma$. The basic unit
of time is $\tau=\sigma(m/\epsilon)^{1/2}$, where $m$ is the mass of
each monomer. The equations of motion are integrated using the LAMMPS
MD simulation package \cite{plimpton95} with a  velocity Verlet
algorithm and a time step $\delta t=0.012 \tau$.  The temperature, $T=\epsilon/k_B$,
was kept constant by weakly coupling the motion of each bead to a heat
bath with a local friction $\Gamma=0.5\tau^{-1}$.

We have studied seven entangled polymer melts of $M$ chains of $N$
beads with $M\times{}N=$ $5000\times{}50$, $2500\times{}100$,
$400\times{}175$, $200\times{}350$, $200\times{}700$,
$400\times{}1000$, and $320\times{}3500$ each at a monomer density
$\rho=0.85\sigma^{-3}$. Using the most refined PPA estimate of the
rheological entanglement length for this model of
$N_e=85\pm7$ \footnote{Ref.~\cite{HoyPRE09} gives a value of $N_e=85$;
we estimate the systematic error related to the use of different PPA algorithms and extraplation schemes to be of the order of $\pm7$.}, the investigated systems span the range
from unentangled ($Z=N/N_e\approx0.6$) to highly entangled
($Z=N/N_e\approx41$).  The Rouse time was previously determined as
$\tau_R=1.5\tau N^2$\cite{kremer90}, entangements effects become
relevant around $\tau_e = \tau_R(N_e)$, and the maximal relaxation
times of entangled systems are expected to be on the order of
$\tau_d^0=\tau_e Z^3=\tau_R Z$.

\begin{figure}[t]
{\centering\resizebox*{1\columnwidth}{!}{
\rotatebox{0}{\includegraphics{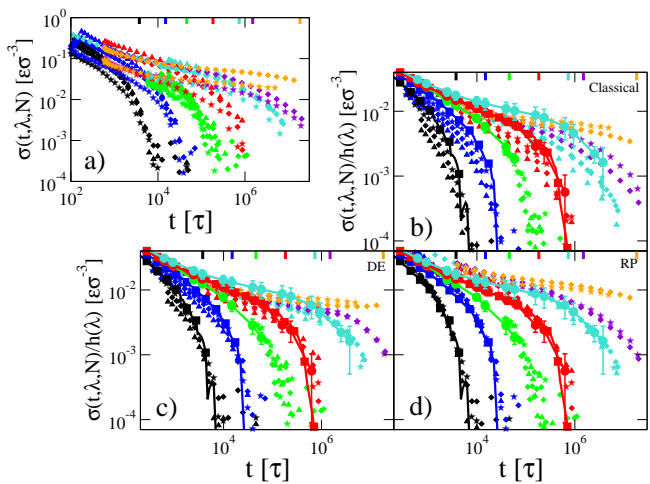}}
} \par{} }
\caption{\label{fig:stress_gt}
  (color online) (a) Normal tensions, $\sigma(\lambda,t)$ for step-strained melts $N=50$ (black), $100$ (blue), $175$ (green), $350$ (red), $700$ (cyan), $1000$ (violet), and $3500$ (orange) and elongation $\lambda=2$ ($\Box$), $3$ ($\diamond$), and $4$ ($\triangle$).
(b-d) Green-Kubo shear relaxation moduli, $G(t)$, (this work: large solid $\bigcirc$ with solid line; Ref.~\cite{LikhtmanMM07}: no symbol, solid line)
 compared to extrapolations, $G(\lambda,t)=\sigma(\lambda,t)/h(\lambda,t)$,
  from the non-linear response using the (b) classical, (c) Doi-Edwards, 
 (d) slip-tube damping function (same symbols as in (a)). Colored ticks indicate the Rouse time of the corresponding 
 systems. 
}
\end{figure}

The melts were generated and equilibrated following the procedure outlined in Auhl
{\it et al.} \cite{AuhlEveraersJCP}. Technically, the largest challenge is  the reliable extraction
of the macroscopic, viscoelastic behavior~\cite{Everaers95,LikhtmanMM07}.
Data were recorded in equilibrium as well as out of equilibrium after a step-strain.
Strained melts were prepared by
subjecting equilibrated initial conformations to rapid ($T_{def}\in [120\tau,36000\tau]$),
 volume-conserving ($\lambda_x \lambda_y \lambda_z\equiv1$),
elongational deformations with ($\lambda_x=\lambda $,
$\lambda_y=$$\lambda_z=$$1/\sqrt{\lambda } $)
\new{ for $\lambda$ ranging from 1.5 to $4.0$ well outside the
linear elastic regime.  Deformations in this range are typical for
many applications of polymeric systems and large enough to generate a measurable elastic
response for the present system sizes.}
In the ideal case, strain should be introduced instantaneously. To check the
dependance of our results on $T_{def}$, we have varied the deformation time in
one case ( $N=700$, $\lambda=3.0$) by a factor of 200. In the following, for deformed
systems, $t=0$ is fixed to the {\em middle} of the deformation period, i.e. data are
recorded for $t\ge T_{def}/2$. To reduce finite-$T_{def}$ artifacts, we typically discard
data from the initial $3T_{def}$.

The longest simulations were run up to $2\times 10^9$ time steps and sufficient
to reach the plateau regime for our longest chains and to completely relax the others
 ($2\times 10^9\times0.012\tau\approx\tau_R(N=4000)\approx\tau_d(Z=12.5)$).
The total numerical effort corresponds to about $5$ million single core CPU hours.
We recorded block-averages of the microscopic
stress tensor $\sigma_{\alpha\beta}(t)=\left\langle \sum_{ij}
F_{ij,\alpha} r_{ij,\beta} \right\rangle/V$ at intervals of $1.2\tau$.
The latter sum is over all pairs $i,j$ of interacting beads, $\alpha$, $\beta$ are Cartesian indices,
and $F$, $r$ and $V$ denotes force, separation and volume, respectively.  Furthermore,
we stored melt conformations at intervals of $120 \tau$ for further analysis of the chain conformations.

Results for the relaxation of the normal tension
$\sigma(t)=\sigma_{xx}-\frac12(\sigma_{yy}+\sigma_{zz})$
are presented in Fig.~\ref{fig:stress_gt}a.
We observe a clear non-Newtonian behavior with a stress relaxation
extending over many orders of magnitude in time after the end
of the deformation period of the sample.
As expected, there is a strong increase of the terminal relaxation
time with chain length and the gradual formation of an
intermediate plateau in the stress relaxation for the longest chains
studied. The maximal relaxation time is independent of the total deformation.

\begin{figure}[t]
{\centering
  \resizebox*{\columnwidth}{!}{\rotatebox{0}{
\includegraphics{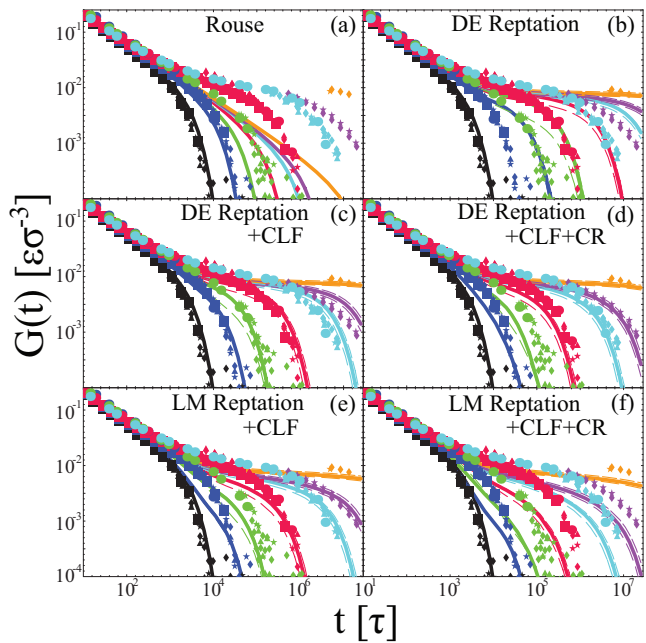}
}}
\par{} }
\caption{\label{fig:GtTheory} (color online) Comparison of the
measured relaxation moduli to the
predictions of various theories.
%(a) the Rouse model~\cite{Rouse_jcp_53,doi86},
%(b) the original Doi-Edwards model accounting only for reptation,
%(c),(d) Doi's modification accounting for the reduction of the tube length due
%to contour length fluctuations without and with the inclusion of constraint release using
%the double reptation approximation,
%(e),(f) the Likhtman-McLeish model including reptation and a more
%sophisticated treatment of contour length fluctuations without and with the inclusion of constraint release using the double reptation approximation.
Symbols and colors for the simulation data as in
Fig.~\protect\ref{fig:stress_gt} with Green-Kubo data shown as large solid $\bigcirc$ and $\Box$. 
Theoretical predictions are shown
as thick lines using the same color code. Thin lines indicate the
uncertainty in the theoretical predictions due
to the uncertainty of
the PPA entanglement length $N_e=85 \pm 7$.}
\end{figure}

\begin{figure}[t]
{\centering\resizebox*{1\columnwidth}{!}{\rotatebox{0}{
\includegraphics{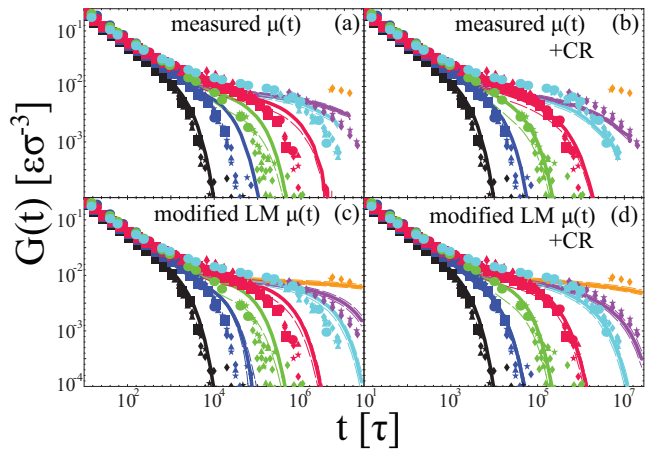}
}} \par{} } \caption{\label{fig:Gt} (color online) 
\new{Comparison of the
measured relaxation moduli to Eq.~(\ref{eq:Gt})
using (a,b) the independently measured autocorrelation function of the 
{\em primitive} chain end-to-end vectors to estimate of the tube memory function,
$\mu(t)$; (c,d) our proposition Eqs.~(\ref{eq:mu_LM_star}-\ref{eq:G_LM_star})
for removing high-frequency modes from the Likhtman-McLeish theory~\protect\cite{LikhtmanMcLeishMM02} of contour length fluctuations with $\alpha=1.7$.
Symbols and colors as in Fig.~\protect\ref{fig:GtTheory}.
}
}
\end{figure}

\new{
Figure~\ref{fig:stress_gt}(b-d) show  comparisons between 
$\sigma(\lambda,t)/h(\lambda,t)$ for the step-strained melts 
to the {\em linear} shear relaxation moduli $G(t)$ obtained
by Likhtman {\it et al.} \cite{LikhtmanMM07} and ourselves
via  the  Green-Kubo
relation $G(t)= V \langle \sigma_{\alpha\beta}(t)
\sigma_{\alpha\beta}(0)\rangle /k_B T$ with $\alpha\neq\beta$
the from stress fluctuations in unstrained, equilibrated melts.
Available expressions for damping functions $h(\lambda)$
are refinements of the stress-strain relation
$h(\lambda)=\lambda^2-\lambda^{-1}$ predicted by
classical rubber elasticity theory (Fig.~\ref{fig:stress_gt}b).
The Doi and Edwards \cite{doi86f}  
damping function 
(Fig.~\ref{fig:stress_gt}c)
includes the dynamics of a uniform chain retraction inside the stretched tube.
The Rubinstein-Panyukov slip-tube damping function \cite{Rubinstein2002},
\begin{math}\label{eq:h_SlipTube}
h(\lambda)=\left(\lambda^2-\lambda^{-1}\right)/\left(0.74\lambda+0.61\lambda^{-1/2}-0.35\right)
\end{math},
accounts for non-affine tube deformations and the asymptotic chain length
redistribution inside the tube (Fig.~\ref{fig:stress_gt}d).
For times $t<\tau_{R}(N)$, the presence of additional relaxation processes
prevents a systematic extraction of $G(t)$ from
normal tensions  measured in the non-linear regime. Empirically,
the time-independent slip-tube expression works surprisingly well.
For times $t>\tau_{R}(N)$, the differences between the slip-tube and
the Doi-Edwards damping functions are small.
Both result in a satisfactory data collapse and good agreement
with the Green-Kubo results, suggesting that
they capture the non-linear effects at large strains with reasonable accuracy.
The step-strain data included in Figs.~\ref{fig:GtTheory} and \ref{fig:Gt} 
correspond to times  $t>0.25\tau_{R}(N)$.
}

In Figure~\ref{fig:GtTheory}a we show a comparison of the simulation results for $G(t)$
to the Rouse model predictions ~\cite{Rouse_jcp_53,doi86}
for unentangled systems. Our results confirm the expectation that the Rouse
model quantitatively describes the chain length independent early-time
stress relaxation with $G(t)\propto t^{-1/2}$ as well as terminal stress relaxation
in systems where the chains are too short to be entangled. 
For longer chains, entanglements start to
affect the behavior beyond a material-specific, characteristic time
$\tau_e\approx 10^4\tau$ with a gradual formation a plateau in
the stress relaxation reached by our longest chain systems with $Z=41$.
For the terminal stress relaxation
of systems with $Z={\cal O}(10)$ we have reliable data extending about one order of magnitude
below the plateau level. This is sufficient to allow for a meaningful
comparison to current theories. In particular, we are not restricted to
comparing the ability of different theories to {\em fit} the data. Rather, we can
carry out  {\em absolute, parameter-free} comparisons using the result $N_e=85\pm7$
\cite{HoyPRE09} of the primitive path analysis and the known Rouse friction of
the model.

Likhtman and McLeish (LM)~\cite{LikhtmanMcLeishMM02} assembled the effects of
(i) early-time Rouse relaxation, (ii) tension equilibration along the contour
of the primitive chains, (iii) reptation, (iv) contour length fluctuations, and (v)
constraint release into a closed functional form,
\begin{eqnarray}\label{eq:Gt}
G(t) &=& \frac{ \rho k_BT}{N} \frac15 \sum_{p=1}^{Z} \big(4\mu(t)
R(t)+e^{- t p^2/\tau_R}\big)\nonumber\\ && + \frac{ \rho k_BT}{N}
\sum_{p=Z+1}^{N} e^{-2 t p^2/\tau_R}
\end{eqnarray}
\noindent where $\mu(t)$ and $R(t)$ account for single- and multi-chain relaxation 
processes of the tube model. In their absence ($\mu(t)=R(t)\equiv1$), the formula describes a crossover from the early time Rouse relaxation
$G(t)\propto t^{-1/2}$ to a plateau $G_N^0= \frac45 \frac{ \rho k_BT}{N_e}$.
The key quantity of the tube model is the single-chain memory function,
$\mu(t)$, for the fraction of the primitive chain which has not escaped from
its original tube after a time $t$. Comparisons to the data
neglecting constraint release ($R(t)\equiv1$) are shown 
for the original Doi-Edwards model accounting
only for reptation (Fig.~\ref{fig:GtTheory}b), Doi's~\cite{doi86} approximate inclusion of the effect of
contour length fluctuations combining reptation dynamics 
with the
maximal relaxation time from~\cite{LikhtmanMcLeishMM02}
(Fig.~\ref{fig:GtTheory}c), and  the
full LM
theory~\cite{LikhtmanMcLeishMM02} of contour length fluctuations and reptation
(Fig.~\ref{fig:GtTheory}e). For the comparisons in
Figs.~\ref{fig:GtTheory}d and f we have included the effect of constraint
release in the double reptation~\cite{descloizeaux88} approximation $R(t)=\mu(t)$.

The overall agreement between our data and the more advanced versions of the
tube model is fairly good. Interestingly, the more sophisticated LM theory seems to work
less well than Doi's approximation when combined with the LM estimate of the
maximal relaxation time.
%theory, which accounts for contour length fluctuations only through
%their reduction of the terminal relaxation time Eq.~(\ref{eq:tau_d}).
Yet, from our rheological data alone,
it is hard to clearly identify the relevance and the quality of the theoretical
description of the various relaxation processes.
For example,  one might (as we believe, erroneously; see below) conclude, that
the double reptation approximation strongly
overestimates the contribution of constraint release to the stress relaxation
(Figs.~\ref{fig:GtTheory}e and f) or that constraint release is inefficient
for $Z<4$ (Figs.~\ref{fig:GtTheory}c and d).
Obviously, fitting the various theories to the data would only obscure their shortcomings.

To draw definite conclusion on how to improve the theories, we
discriminate between three possible sources
of error:
(i) the functional form of Eq.~(\ref{eq:Gt}),
(ii) the treatment of reptation and contour length fluctuations
underlying the single-chain memory function $\mu(t)$, and
(iii) the treatment of the multi-chain effect of constraint release via
the double reptation approximation, $R(t)=\mu(t)$.
\new{
For long chains and under the assumption that the escaped chain sections equilibrate completely,
$\mu(t)$ equals the autocorrelation function of the chain end-to-end vectors~\cite{mcleish02},
For shorter chains, it is more suitable to consider the end-point motion of the
{\em primitive} chains, defined as the average of the chain conformation over
a period of $\tau_e$ \cite{Read_mm_08}. 
This correlation function is easily accessible from our equilibrium simulations
and is {\em not} affected by constraint release~\cite{mcleish02}.
The comparison between the
measured relaxation moduli and those predicted from Eq.~(\ref{eq:Gt})
using the measured $\mu(t)$ together with $R(t)\equiv1$ and $R(t)=\mu(t)$ 
 is shown in Figs.~\ref{fig:Gt} a and b respectively.
For the full theory the agreement is excellent, supporting the utility of both  the
Likhtman-McLeish functional form of the shear relaxation modulus and of the
double reptation approximation for constraint release. 
%More elaborate schemes~\cite{RubinsteinColbyJCP88} for predicting $R\left[\mu(t)\right]$ should lead to even better agreement.
%We note, that we have not measured the decay of G(t) to sufficiently small values to observe the
%predicted~\cite{RubinsteinColbyJCP88,LikhtmanMcLeishMM02} long time
%tail in the terminal decay.
%It is, however, tempting to ascribe the slight overestimation of the stress relaxation
%for the longest chains to the neglect of this effect by the double reptation approximation.
}

The shortcomings of the LM description apparent in Fig.~\ref{fig:GtTheory}f and in the
rheological study by Liu {\it et al.}~\cite{BaillyPRL06}  must thus be
related to the central part of their theory, the estimation of the time dependence
of $\mu(t)$ under the combined influence
of reptation and contour length fluctuations. 
\new{
A possible explanation is a double-counting of the effect of short-wavelength ($p>Z$) modes
in the Rouse relaxation part of Eq.~(\ref{eq:Gt}) and in $\mu(t)$. LM
extrapolated $\mu(t)$ to the continuum limit, resulting in a decay on
time scales $t<\tau_e$, where the motion of the {\em primitive} chain should be negligible.
To correct for this, we have removed from the CLF part of $\mu(t)$ 
the contribution of modes with a relaxation time shorter 
than $\alpha^4 \tau_e$:
%
%{\bf Likhtman's original formulas}
%\begin{eqnarray}
%\mu(t) &=&
%\frac{8\widetilde{G}_f}{\pi^2}\sum_{p=1,\mathrm{odd}}^{p^*}{\frac{1}{p^2}\exp{\left(-\frac{t
%p^2}{\tau_{df}}\right)}}\nonumber\\
%&&+\int_{\epsilon^*}^{\infty}\frac{0.306}{Z \tau^{1/4}_e
%\epsilon^{5/4}}\exp(-\epsilon t)d\epsilon
%\end{eqnarray}
%
%\begin{eqnarray}
%\epsilon^*=\frac{1}{\tau_e
%Z^4}\left(\frac{4\times0.306}{1-{\frac{8\widetilde{G}_f}{\pi^2}}{\sum_{p=1,\mathrm{odd}}^{p^*}\frac{1}{p^2}}}\right)^4
%\end{eqnarray}
%
%\begin{eqnarray}
%\frac{\tau_{df}(Z)}{\tau_R}=3Z\left(1-\frac{2\times1.69}{\sqrt{Z}}+\frac{4.17}{Z}-\frac{1.55}{Z^{3/2}}\right)
%\end{eqnarray}
%
%
%\begin{eqnarray}
%\widetilde{G}_f=1-\frac{1.69}{\sqrt{Z}}+\frac{2.0}{Z}-\frac{1.24}{Z^{3/2}}
%\end{eqnarray}
%
%\begin{eqnarray}
%p^*=\sqrt{Z/10}
%\end{eqnarray}
%
\begin{eqnarray}
\label{eq:mu_LM_star}
\mu(t) &=&\left(1+\frac{1.22\alpha}Z\right) \left(
\frac{8\widetilde{G}_f}{\pi^2}\sum_{p=1,\mathrm{odd}}^{p^*}{\frac{1}{p^2}\exp{\left(-\frac{t
p^2}{\tau_{df}}\right)}}    \right.\nonumber\\
&&\left. +\int_{\epsilon^*}^{1/(\alpha^4 \tau_e)}\frac{0.306}{Z \tau^{1/4}_e
\epsilon^{5/4}}\exp(-\epsilon t)d\epsilon\right)\\
\label{eq:tau_max_LM_star}
\frac{\tau_{df}(Z)}{\tau_R}&=&3Z\Big(1-\frac{2\times1.69}{\sqrt{Z}}+\frac{4.17+1.22\alpha}{Z}\nonumber\\
&&-\frac{1.55+2.69\alpha}{Z^{3/2}}\Big)\\
\label{eq:G_LM_star}
\widetilde{G}_f&=&1-\frac{1.69}{\sqrt{Z}}+\frac{2.0}{Z}-\frac{1.24-1.03\alpha}{Z^{3/2}}
%-\frac{2.448\alpha}{Z^2}
\end{eqnarray}
where $p^*=\sqrt{Z/10}$ and $\epsilon^*$ are defined as in the original LM theory,
which is recovered in the $\alpha\equiv0$ limit of the above expressions. Fig.~\ref{fig:Gt}d shows
that we obtain significantly improved agreement between theory and our data
for values $\alpha={\cal O}(1)$. Interestingly, this corresponds to a constant offset
of $Z=Z_{LM}-\int_{1/(\alpha^4 \tau_e)}^\infty \frac{0.306}{Z \tau^{1/4}_e
\epsilon^{5/4}}\exp(-\epsilon t)d\epsilon=Z_{LM}-1.22\alpha$. 
This view is in qualitative agreement with 
arguments put forward by van Ruymbeke {\em et al.}~ \cite{Bailly_mm_10} to consider, within the original 
LM theory, chains with virtual extensions of length $~N_e$ resulting in an increase
the relaxation time of the outermost ``real'' chain segment to $\tau_e$.
%We conclude 
% thus reconciling the description
%of rheological data and of neutron-spin echo results on the conformational relaxation~\cite{Wischnewski_prl_02} 
%
%
% since
%the inclusion of contour length fluctuations is essential for obtaining a coherent description
%of neutron-spin echo results on the conformational relaxation~\cite{Wischnewski_prl_02}.
%
}%new

To summarize, we have presented an extensive set of simulation results for the
equilibrium and relaxation dynamics of entangled model polymer melts.
%We have shown that the use of the stress-strain relation of
%the slip-tube model~\cite{Rubinstein2002} as damping function
%results in excellent agreement between equilibrium and non-equilibrium
%measurements of $G(t)$.
In particular, we explored $G(t)$ into
the plateau regime for chains with $Z=41$ and into the terminal relaxation
regime for $Z\le 10$ and compared our
data to predictions of different versions of the tube model. These comparisons did
not involve any free parameters, since the entanglement length was
determined independently via a topological analysis~\cite{RE_PPA_sci_04,HoyPRE09}.
\new{
We find excellent agreement for the Liktman-McLeish theory using a corrected
tube memory function and the double reptation approximation for constraint release},
demonstrating that the primitive path analysis  of the microscopic
{\em structure} endows the tube model with predictive power
for {\em dynamical processes}.
%Our results illustrate that with the presently accessible
%chain lengths and time scales and with the simultaneous access to
%stress and conformational relaxation, computer simulations
%are now in a position to subject tube models to detailed tests to contribute to the 
%systematic development of improved theories.
The use of more elaborate schemes~\cite{RubinsteinColbyJCP88} for treating constraint release
and predicting the function $R\left[\mu(t)\right]$ should lead to even better agreement.
%Longer runs should enable us to measure
%the decay of G(t) to sufficiently small values to observe the
%predicted~\cite{RubinsteinColbyJCP88,LikhtmanMcLeishMM02} long time
%tail in the terminal decay. At present, it is 
% tempting to ascribe the slight overestimation of the stress relaxation
%for the longest chains to the neglect of this effect by the double reptation approximation.

\begin{acknowledgments} We thank the New Mexico Computing Application Center NMCAC
for generous allocation of computer time and A.
Likhtman and S.K. Sukumaran for their Green-Kubo data.
%\new{and D. Read and A. Likhtman for useful discussions during the 2010 APS March Meeting}
CS acknowledges financial
support from the Danish Natural Sciences Research Council through a
Steno Research Assistant Professor fellowship.
JXH is supported by the EC through the Marie Curie EST {\it Eurosim} Project No. MEST-CT-2005-
020491. RE acknowledges a
chair of excellence grant from the Agence Nationale de Recherche
(France). Sandia is a multiprogram
laboratory operated by Sandia Corporation, a Lockheed Martin Company,
for the U.S. Department of Energy under Contract No.
DE-AC04-94AL85000. \end{acknowledgments}

\end{document}